\documentclass[conf]{new-aiaa} 

\usepackage[utf8]{inputenc}
\usepackage{textcomp}
\usepackage{tikz}
\usetikzlibrary{shapes,arrows}
\usepackage{graphicx}
\usepackage{amsmath}
\usepackage[version=4]{mhchem}
\usepackage{siunitx}
\usepackage{combelow}
\usepackage{multirow}
\usepackage{multicol}
\usepackage{xcolor}
\usepackage{amsfonts}
\usepackage{todonotes}
\usepackage{mathtools}
\usepackage{arydshln}
\usepackage{algorithm}
\usepackage{algpseudocode}
\usepackage{nomencl}
\usepackage{fancyhdr}
\usepackage{longtable,tabularx}
\usepackage{upgreek}
\usepackage{listings}
\usepackage{bm}

\newtheorem{remark}{Remark}

\definecolor{antiquebrass}{rgb}{0.8, 0.58, 0.46}
\definecolor{carminepink}{rgb}{0.92, 0.3, 0.26}
\definecolor{cobalt}{rgb}{0.0, 0.28, 0.67}
\definecolor{cerulean}{rgb}{0.0, 0.48, 0.65}

\definecolor{codegreen}{rgb}{0,0.6,0}
\definecolor{codegray}{rgb}{0.5,0.5,0.5}
\definecolor{codepurple}{rgb}{0.58,0,0.82}
\definecolor{backcolour}{rgb}{0.95,0.95,0.92}

\lstdefinestyle{mystyle}{
    backgroundcolor=\color{backcolour},   
    commentstyle=\color{codegreen},
    keywordstyle=\color{magenta},
    numberstyle=\tiny\color{codegray},
    stringstyle=\color{codepurple},
    basicstyle=\ttfamily\footnotesize,
    breakatwhitespace=false,         
    breaklines=true,                 
    captionpos=b,                    
    keepspaces=true,                 
    numbers=left,                    
    numbersep=5pt,                  
    showspaces=false,                
    showstringspaces=false,
    showtabs=false,                  
    tabsize=2
}

\newcount\Comments
\newcommand{\kibitz}[2]{\ifnum\Comments=1\textcolor{#1}{#2}\fi}

\title{A parallel implementation of reduced-order modeling of large-scale systems}

\author{Ionu\cb{t}-Gabriel Farca\cb{s} \footnote{Assistant Professor, Department of Mathematics, farcasi@vt.edu, AIAA Member.}\footnote{Oden Institute for Computational Engineering and Sciences, ionut.farcas@austin.utexas.edu}}
\affil{Virginia Tech, Blacksburg VA, 24061}
\affil{The University of Texas at Austin, Austin TX, 78712}
\author{Rayomand P. Gundevia  \footnote{Research Scientist, rayomand.gundevia.ctr@afrl.af.mil, AIAA Member.}}
\affil{Amentum, Edwards AFB CA 93524}
\author{Ramakanth Munipalli\footnote{Senior Aerospace Research Engineer, Combustion Devices, ramakanth.munipalli@us.af.mil, AIAA Senior Member.}}
\affil{Air Force Research Laboratory, Edwards AFB CA 93524}
\author{Karen E.~Willcox\footnote{Director, Oden Institute for Computational Engineering and Sciences, kwillcox@oden.utexas.edu, AIAA Fellow.}}
\affil{The University of Texas at Austin, Austin TX, 78712}

\begin{document}

\maketitle

\begin{abstract}
Motivated by the large-scale nature of modern aerospace engineering simulations, this paper presents a detailed description of distributed Operator Inference (dOpInf), a recently developed parallel algorithm designed to efficiently construct physics-based reduced-order models (ROMs) for problems with large state dimensions.
One such example is the simulation of rotating detonation rocket engines, where snapshot data generated by high-fidelity large-eddy simulations have many millions of degrees of freedom.
dOpInf enables, via distributed computing, the efficient processing of datasets with state dimensions that are too large to process on a single computer, and the learning of structured physics-based ROMs that approximate the dynamical systems underlying those datasets.
All elements of dOpInf are scalable, leading to a fully parallelized reduced modeling approach that can scale to the thousands of processors available on leadership high-performance computing platforms.
The resulting ROMs are computationally cheap, making them ideal for key engineering tasks such as design space exploration, risk assessment, and uncertainty quantification.
To illustrate the practical application of dOpInf, we provide a step-by-step tutorial using a 2D Navier-Stokes flow over a step scenario as a case study.
This tutorial guides users through the implementation process, making dOpInf accessible for integration into complex aerospace engineering simulations.
\end{abstract}

\section{Introduction}

The ever-increasing complexity of aerospace engineering simulations, particularly for applications like rotating detonation rocket engines (RDREs), demands innovative approaches for constructing reduced-order models (ROMs) to provide computationally cheap yet sufficiently accurate surrogates for the high-fidelity model.
ROMs are crucial for enabling engineering tasks such as design exploration, risk assessment, and uncertainty quantification that are infeasible in terms of high-fidelity models, which remain computationally expensive to simulate even on powerful supercomputers.
This paper provides a detailed description, in a tutorial style, of a recently developed distributed Operator Inference (dOpInf) algorithm~\cite{farcas2024distributedcomputingphysicsbaseddatadriven} for constructing physics-based ROMs efficiently and scalably on parallel machines.
Our goal is to provide clear, step-by-step instructions on implementing dOpInf and to facilitate its integration into existing aerospace engineering workflows.

dOpInf provides a complete distributed data-driven reduced modeling procedure to tackle problems with large state dimensions, enabling a fast and scalable construction of structured, physics-based ROMs that approximate the dynamical systems underlying those datasets~\cite{PW16, KPW24}.
A key factor in its scalability is an efficient data dimensionality reduction based on the proper orthogonal decomposition (POD) method of snapshots~\cite{Si87}. 
We use the method of snapshots to devise a technique that bypasses the need to compute the reduced basis to perform the dimensionality reduction, thereby improving performance.
In addition, dOpInf primarily relies on standard dense numerical linear algebra operations, such as matrix-matrix multiplications and eigenvalue decompositions. 
These operations are fundamental to many scientific computing tasks and are efficiently implemented in most scientific computing libraries.

The development of algorithms and software tools to enable model reduction in large-scale applications is an active area of research.
Reference~\cite{Be06} proposed an approximate iterative parallel algorithm and Ref.~\cite{WMI16} formulated an approximate distributed approach based on the method of snapshots for approximating POD.
The recent work~\cite{Ke23} formulated a high-performance SVD solver for computing partial SVD spectra. 
Exploiting the fact that in large-scale applications the data matrices containing the training snapshots are tall-and-skinny (TS), that is, the snapshot dimension is significantly larger than the number of snapshots, Ref.~\cite{De12} formulated a parallel QR-decomposition (TSQR) algorithm which was later used by Ref.~\cite{Be13} for a MapReduce infrastructure, and leveraged in Ref.~\cite{Sayadi2016} for performing Dynamic Mode Decomposition (DMD)~\cite{Sc10, Ku16, Tu14} efficiently in parallel.
An implementation of TSQR is also provided in the Trilinos software framework~\cite{trilinos}.
A complementary perspective is offered by streaming or sequential methods, which bypass the need for expensive disk input/output operations as well as storing and loading large datasets once they have been processed.
Reference~\cite{LL98} proposed an incremental procedure for computing the POD from streaming data.
This procedure was leveraged in Ref.~\cite{schwerdtner2024onlinelearningquadraticmanifolds}, which formulated a sequential, greedy approach for constructing approximations based on quadratic manifolds.
Software tools for model reduction include \texttt{libROM}~\cite{libROM}, a \texttt{C++} library for data-driven physical simulation methods, \texttt{Pressio}~\cite{rizzi2021pressioenablingprojectionbasedmodel}, an open-source project aimed at providing intrusive model reduction capabilities to large-scale application codes, \texttt{PyMOR}~\cite{MRS16}, a \texttt{Python} library for model order reduction algorithms, in particular reduced basis methods, and the recently developed \texttt{PySPOD} library~\cite{Ro24} for parallel spectral POD.

This paper focuses on dOpInf ROMs utilized for predictions beyond a training time horizon, but our ideas are applicable to other data-driven reduced modeling approaches such as DMD and quadratic manifolds~\cite{GWW23, barnett2022quadratic}, and to parametric ROMs that embed parametric dependence using the strategies surveyed in~\cite{BGW15}.
The full code is available at \url{https://github.com/ionutfarcas/distributed_Operator_Inference}.
The repository contains a \texttt{Jupyter Notebook} that offers a step-by-step tutorial on applying the dOpInf algorithm to a 2D Navier-Stokes problem.
This tutorial is intended to provide users with a comprehensive understanding of the algorithm's mechanics.
For details into the scalability of dOpInf in a large-scale RDRE scenario, we refer the reader to Ref.~\cite{farcas2024distributedcomputingphysicsbaseddatadriven}.
The repository also includes the training dataset and additional \texttt{Python} scripts for simulating the high-fidelity model, extracting indices for specific probe locations for comparing the reference with the reduced solutions, reference high-fidelity data for postprocessing purposes, and a reference serial implementation of OpInf.

The remainder of this paper is organized as follows.
Section~\ref{sec:preliminaries} summarizes the high-fidelity simulation framework and presents a 2D Navier-Stokes flow over a step scenario as a case study to demonstrate our distributed algorithm.
Section~\ref{sec:dOpInf_tutorial} provides a detailed, tutorial-style explanation of the dOpInf algorithm. 
For each step, we discuss the underlying mathematical and computational concepts and provide \texttt{Python} code snippets to illustrate their practical application to the 2D Navier-Stokes example.
We discuss the results in Sec.~\ref{sec:results} and summarize the paper in Sec.~\ref{sec:conclusion}.

\section{Preliminaries} \label{sec:preliminaries}

Section~\ref{subsec:sec_2_preliminaries} summarizes the setup for high-fidelity simulations and Sec.~\ref{subsec:sec_2_model_problem} presents the 2D transient flow past a circular cylinder scenario used to demonstrate the practical implementation of our dOpInf algorithm.

\subsection{Setup for high-fidelity simulations} \label{subsec:sec_2_preliminaries}
Consider a physical process of interest whose temporal dynamics over the time horizon $[t_{\textrm{init}}, t_{\textrm{final}}]$, where $t_{\textrm{init}}$ denotes the initial time and $t_{\textrm{final}}$ the final time, are described by the dynamical system 
\begin{equation} \label{eq:FOM_general}
    \dot{\mathbf{s}} = \mathbf{f}(t, \mathbf{s}), \quad \mathbf{s}(t_{\textrm{init}})=\mathbf{s}_{\mathrm{init}},
\end{equation}
where $\mathbf{s}(t) \in \mathbb{R}^{n}$ denotes the vector of state variables at time $t \in [t_{\textrm{init}}, t_{\textrm{final}}]$, $n$ is the large dimension of the state space, $\mathbf{s}_{\mathrm{init}}$ is a specified initial condition, and $\mathbf{f} : [t_{\textrm{init}}, t_{\textrm{final}}] \times \mathbb{R}^{n} \rightarrow \mathbb{R}^{n}$ is a nonlinear function that defines the time evolution of the high-dimensional state $\mathbf{s}$.

Our starting point is a given set of training data representing observations of the system governed by \eqref{eq:FOM_general}. 
These data might be generated through experiments, simulations, or a combination of both. 
In many cases, the data come from simulations that solve the physical equations governing the system's behavior. 
In aerospace engineering applications, the governing equations are typically PDEs that represent physical conservation laws and constitutive relationships.
When discretized in space, these equations take the form of \eqref{eq:FOM_general}.
In large-scale scenarios, $n$ is generally very large (e.g., in $\mathcal{O}(10^6)$--$\mathcal{O}(10^9)$) because it scales with the dimension of the PDE spatial discretization.

The training data set comprises $n_t$ state vectors at $n_t$ time instants over a training horizon $[t_{\textrm{init}}, t_{\textrm{train}}]$ with $t_{\textrm{train}} < t_{\textrm{final}}$.
The state solution at time $t_k \in [t_{\textrm{init}}, t_{\textrm{train}}]$ is referred to as the $k$th snapshot and is denoted by $\mathbf{s}_k$.
The $n_t$ snapshots are then collected into a snapshot matrix $\mathbf{S} \in \mathbb{R}^{n \times n_t}$ with $\mathbf{s}_k$ as its $k$th column:
\begin{equation*}
    \mathbf{S} =
     \begin{bmatrix}
\vert & \vert & & \vert\\
     \mathbf{s}_1 &
     \mathbf{s}_2 &
     \ldots &
     \mathbf{s}_{n_t}\\
     \vert & \vert & & \vert
     \end{bmatrix}.
\end{equation*}
This paper focuses on constructing predictive physics-based ROMs over the target time horizon $[t_{\textrm{init}}, t_{\textrm{final}}]$ for systems with large dimensions $n$ for which standard, serial data-driven methods are impractical due to their memory and computational requirements.

\subsection{Flow over a cylinder in a two-dimensional spatial domain} \label{subsec:sec_2_model_problem}
\begin{figure}[htp]
\centering
\includegraphics[width=0.7\textwidth]{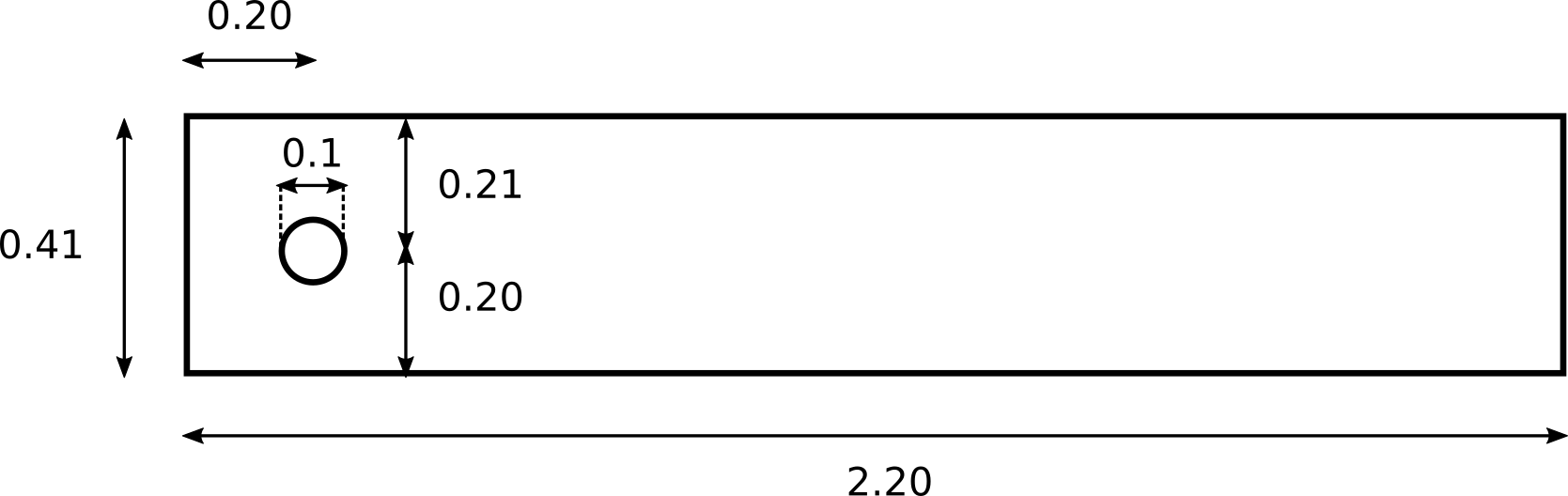}
\caption{Physical domain of the considered 2D transient flow past a circular cylinder scenario.}
\label{fig:2D_bechmark_domain}
\end{figure}

To demonstrate the practical implementation of the dOpInf algorithm, we consider the canonical problem of two-dimensional transient flow past a circular cylinder, a widely-used benchmark in computational fluid dynamics and reduced-order modeling.
The fluid flow is governed by the 2D incompressible Navier-Stokes equations
\begin{align} \label{eq:PDE_Navier_Stokes}
\partial_t \mathbf{u} + \nabla \cdot (\mathbf{u} \otimes \mathbf{u}) & = \nabla p + Re^{-1}\Delta \mathbf{u} \\
\nabla \cdot \mathbf{u} & = 0,
\end{align}
where $p \in \mathbb{R}$ denotes the pressure, $\mathbf{u} = (u_x, u_y)^\top \in \mathbb{R}^2$ denotes the $x$ and $y$ components of the velocity vector, and $Re$ denotes the Reynolds number.
Figure~\ref{fig:2D_bechmark_domain} plots the physical domain.
The problem setup, geometry, and parameterization are based on the DFG 2D-3 benchmark in the FeatFlow suite\footnote{\url{https://wwwold.mathematik.tu-dortmund.de/~featflow/en/benchmarks/cfdbenchmarking/flow/dfg_benchmark2_re100.html}}.

We solve the Navier Stokes equations~\eqref{eq:PDE_Navier_Stokes} numerically in double precision using finite elements with the \texttt{FEniCS} software package.
This implementation, provided in our repository, closely follows the reference implementation found at \url{https://fenicsproject.org/pub/tutorial/html/._ftut1009.html}.
It uses piecewise quadratic finite elements for the velocity and piecewise linear elements for the pressure.
The resulting system of equations is large and sparse, making direct solvers impractical.  
We employ iterative Krylov subspace methods with preconditioning for improved performance. 
Specifically, we utilize the biconjugate gradient stabilized (BiCGstab) method and the conjugate gradient (CG) method.
The setup for the DFG 2D-3 benchmark uses $Re = 100$, which represents a value that is above the critical Reynolds number for the onset of the two-dimensional vortex shedding.

We employ a finite element mesh with $n_x = 146,339$ degrees of freedom.
The discretized governing equations are integrated over the time interval $[0, 10]$ seconds in increments of $\Delta t = 2.5 \times 10^{-5}$ seconds.
This amounts to a total of $40,000$ snapshots.
In the model reduction experiments that follow, we simplify the model by omitting the pressure term. 
This simplification is justified by the fact that, for both transient and periodic flow regimes, the pressure can be implicitly determined from the velocity field through the incompressibility constraint.
The target time horizon is $[4, 10]$ seconds, corresponding to the periodic regime.
The training data are collected over $[4, 7]$ seconds, whereas the remainder of the target time horizon is used for predictions beyond training.
\texttt{FEniCS} offers functionality to save simulation data do disk in HDF5 format\footnote{\url{https://www.hdfgroup.org/solutions/hdf5/}}, which we also leverage for our numerical experiments. 
Nonetheless, storing the full velocity fields over the training horizon would require approximately $28$ GB of storage, exceeding the capabilities of a typical personal computer and hindering the accessibility of our dOpInf tutorial.
To address this challenge, we downsample the training data by a factor of $20$, reducing its storage footprint to a manageable size of around $1.4$ GB for $600$ downsampled snapshots.
The training data were saved to disk in a single HDF5 file.
This file contains two datasets, $u\_x$ and $u\_y$, for the two downsampled velocity components of dimension $146,339 \times 600$.

\section{Distributed Operator Inference for physics-based data-driven reduced modeling of large-scale systems: a step-by-step tutorial} \label{sec:dOpInf_tutorial}

This section presents, in a tutorial style, our recently developed dOpInf algorithm~\cite{farcas2024distributedcomputingphysicsbaseddatadriven}.
We start by summarizing the setup for parallel computations in Sec.~\ref{subsec:dOpInf_setup}.
The following five sections describe the steps to implement the dOpInf algorithm in practice. 
Section~\ref{subsec:dOpInf_step_I} presents the distributed training data loading step.
Section~\ref{subsec:dOpInf_step_II} focuses on training data transformations, followed by parallel data dimensionality reduction in Sec.~\ref{subsec:dOpInf_step_III}.
We then use distributed computing for learning the reduced model operators, presented in Sec.~\ref{subsec:dOpInf_step_IV}.
Section~\ref{subsec:dOpInf_step_V} shows how the obtained reduced solution can be postprocessed in parallel.
For all steps, we provide code snippets in \texttt{Python} to showcase their practical implementation for the 2D Navier-Stokes example summarized in Sec.~\ref{subsec:sec_2_model_problem}.
Our main goal is to guide users on practical details and facilitate a seamless integration with complex aerospace engineering applications.

\subsection{Setup for parallel computations} \label{subsec:dOpInf_setup}

Let $p \in \mathbb{N}_{\geq 2}$ denote the number of compute cores to perform dOpInf.
This paper focuses on a dOpInf implementation using the Message Passing Interface (MPI) utilizing a single group with $p$ MPI ranks (i.e., with one rank per compute core) on a CPU-based parallel machine.
For convenience, we use the notation $i = 0, 1, \dots, p-1$ to denote both the $i$th core and its corresponding MPI rank.  
We note, nonetheless, that our implementation can be extended, for example, to using $p$ accelerator cores if an accelerator-based machine is employed (e.g., GPUs or tensor processing units~\cite{Le22}).

In the following, we present in detail the steps to perform the dOpInf algorithm.
For each step, we first discuss the mathematical and computational aspects followed by its distributed implementation.
We provide code snippets in \texttt{Python} with detailed comments to showcase the implementation of each step in the 2D Navier-Stokes example presented in Sec.~\ref{subsec:sec_2_model_problem}.
Our distributed implementation is done using the MPI for \texttt{Python} library \texttt{mpi4py}\footnote{\url{https://mpi4py.readthedocs.io/en/stable/}}.

\subsection{Step I: Parallel training data loading} \label{subsec:dOpInf_step_I}
We assume that the training dataset, generated through experiments, simulations, or a combination thereof, is readily available.  
In the first step, we load it in parallel from disk.
Efficient parallel loading of datasets with large state dimension $n$ is essential for performance. 
We recommend using file formats like HDF, which support scalable parallel I/O operations through features like independent data access and parallel reading and writing operations.
\begin{remark}
    To ensure a scalable reading of the training dataset across $p$ compute cores, a deep understanding of the underlying file system is crucial.
    Nonetheless, the reading scalability usually deteriorates as $p$ increases when the entire dataset is saved in a single file.
    In this situation, we recommend file partitioning, that is, the dataset is divided into multiple files, allowing for parallel reading operations across different processes, or distributed reading and broadcasting, in which, for example, the first rank reads the data and broadcasts it to all other ranks.
\end{remark}

\subsubsection{Mathematical and computational aspects}

We must first choose a strategy for distributing the full training dataset $\mathbf{S} \in \mathbb{{R}}^{n \times n_t}$ into $p$ non-overlapping blocks $\mathbf{S}_i \in \mathbb{R}^{n_i \times n_t}$ for $i = 0, 1, \dots, p-1$ such that $\sum_{i = 0}^{p-1} n_i = n$. 
The splitting strategy is user-specified and depends on the target parallel architecture.
For example, a straightforward strategy suitable for homogeneous architectures is to use blocks of size $n_i = \lfloor n/p \rfloor$, and when $p$ does not divide $n$ exactly, to further distribute the remaining $n - \lfloor n/p \rfloor$ rows among the $p$ cores.
Each compute core then loads $\mathbf{S}_i$ into its corresponding main memory.

\subsubsection{Parallel implementation details}
We start by importing necessary libraries, defining key variables, and initializing the standard MPI communicator:
\begin{lstlisting}[language=Python]
import numpy as np
from mpi4py import MPI
import h5py as h5

######## INITIALIZATION ########
# DoF setup
ns 	= 2
n 	= 292678 
nx  = int(n/ns)

# number of training snapshots
nt = 600

# state variable names
state_variables = ['u_x', 'u_y']

# path to the HDF5 file containing the training snapshots
H5_training_snapshots = 'navier_stokes_benchmark/velocity_training_snapshots.h5'

# number of time instants over the time domain of interest (training + prediction)
nt_p = 1200

# MPI initialization
comm = MPI.COMM_WORLD
rank = comm.Get_rank()
size = comm.Get_size()
###### INITIALIZATION END ######
\end{lstlisting}
For all linear algebra operations, we will use the \texttt{numpy} library.\footnote{\url{https://numpy.org/}}.
All MPI-related functionality is provided by \texttt{mpi4py}.
Since the snapshot data for the 2D Navier-Stokes example was saved to disk in HDF5 format, we use the functionality provided by the \texttt{h5py}\footnote{\url{https://www.h5py.org/}} library for loading it in parallel. 

We proceed by defining several key variables, such as the number of physical state variables (\texttt{ns = 2}), the total snapshot dimension (\texttt{n = 292678}), the number of spatial degrees of freedom (\texttt{nx = int(n/ns)}), and the number of training snapshots (\texttt{nt = 600}).
We also define a list, \texttt{state\_variables = ['u\_x', 'u\_y']}, containing the names of the two state variables saved in the HDF5 file, \texttt{H5\_training\_snapshots}, which stores the training snapshots.
Finally, we define the number of time instants over the target time horizon $[4, 10]$ seconds, \texttt{nt\_p = 1200}.

We then initialize the default MPI intra-communicator, MPI COMM WORLD. 
The rank of each compute core is given by the variable \texttt{rank} ranging from $0$ to $p-1$, whereas the total size of the communicator ($p$ in our case) is given by the variable \texttt{size}.
\begin{remark}
Recall that MPI programs are inherently parallel, with each rank executing its own copy of the program.
In dOpinf, each rank processes its assigned block $\mathbf{S}_i \in \mathbb{R}^{n_i \times n_t}$ of the full training dataset and when required, results are aggregated to compute global values. 
To achieve scalability, it is crucial to minimize communication overhead and leverage non-blocking communication whenever possible.
For instance, using asynchronous communication for sending and receiving data can help overlap computation with communication.
\end{remark}

In the next step we load the training data in parallel, as shown by the code below.
We first define a function \texttt{distribute\_nx} to distribute the number of spatial degrees of freedom $n_x$ into $p$ independent components $n_{x, i}$ such that $\sum_{i=0}^{p-1} n_{x, i} = 146,339$.
Moreover, $n_i = 2 n_{x, i}$ and $\sum_{i=0}^{p-1} n_{i} = 292,678$.
Note that this corresponds to decomposing the spatial domain into $p$ non-overlapping subdomains.
Therefore, each compute core contains the data for each state variable over a portion of the physical domain.
This splitting scheme allows to efficiently perform data transformations in parallel; we further discuss this point in Sec.~\ref{subsec:dOpInf_step_II}.
Each rank then reads the two velocity components corresponding to its subdomain into a snapshot matrix denoted by \texttt{Q\_rank} of size $2 n_{x, i} \times n_t = n_{i} \times n_t$. 

\begin{lstlisting}[language=Python, firstnumber=28]
######## STEP I: DISTRIBUTED TRAINING DATA LOADING ########
def distribute_nx(rank, nx, size):
	"""
 	distribute_nx distributes the spatial DoF nx into chunks of size nx_i such that 
 	\sum_{i=0}^{p-1} nx_i = nx where p is the number of used compute cores

 	:rank: 	the MPI rank i = 0, 1, ... , p-1 that will execute this function
 	:n_x: 	number of DoF used for spatial discretization
 	:size: 	size of the MPI communicator (p in our case)
 	
 	:return: the start and end index of the local DoF, and the number of local DoF for each rank
 	"""

	nx_i_equal = int(nx/size)

	nx_i_start = rank * nx_i_equal
	nx_i_end   = (rank + 1) * nx_i_equal

	if rank == size - 1 and nx_i_end != nx:
		nx_i_end += nx - size*nx_i_equal

	nx_i = nx_i_end - nx_i_start

	return nx_i_start, nx_i_end, nx_i

# the start and end indices, and the total number of snapshots for each MPI rank
nx_i_start, nx_i_end, nx_i = distribute_nx(rank, nx, size)
        
# allocate memory for the snapshot data corresponding to each MPI rank
# the full snapshot data has been saved to disk in HDF5 format
Q_rank = np.zeros((ns * nx_i, nt))
with h5.File(H5_training_snapshots, 'r') as file:

    for j in range(ns):
        Q_rank[j*nx_i : (j + 1)*nx_i, :] = \
		          file[state_variables[j]][nx_i_start : nx_i_end, :]

file.close()
#################### STEP I END ###########################
\end{lstlisting}

\subsection{Step II: Parallel training data manipulations}  \label{subsec:dOpInf_step_II}
In the next step, we perform any necessary data manipulations, which can vary depending on the specific problem. 
This might include variable transformations for scenarios with non-polynomial governing equations or scaling transformations for those involving multiple states with varying scales.

\subsubsection{Mathematical and computational aspects}
One example of data transformation specific to OpInf is lifting, used to expose polynomial structure in the lifted governing equations for systems with non-polynomial structure~\cite{Qi20}.
If the governing equations are polynomial in the underlying state variables, as in the case of the considered 2D Navier-Stokes example, lifting transformations are unnecessary.
Another commonly used data transformation, especially in problems with multiple state variables, is data centering.
One strategy is to center the (possibly lifted) snapshots variables-by-variable by their respective temporal means over the training horizon.
When the state variables have significantly different scales (like pressure and chemical mass fractions in reacting flows), scaling is also crucial to prevent undesirable bias towards variables with larger values in the corresponding POD bases.
Scaling can be done, for example, with respect to the absolute maximum value of each centered variable, which ensures that the scaled variables do not exceed $[-1, 1]$; other strategies such as using the standard deviation of the centered variables can also be employed.
In all our experiments with complex reactive flow simulations~\cite{Sw20, MHW21, QFW21}, for example, centering and scaling were essential for constructing predictive OpInf ROMs.

If data transformations are used, we denote by $\mathbf{Q}_i \in \mathbb{R}^{m_i \times n_t}$ the \emph{transformed} snapshot data on each compute core, with $\sum_{i=0}^{p-1} m_i = m \geq n$ denoting the transformed snapshot dimension.
Note that $m$ exceeds $n$ when the number of lifted variables exceeds the number of original state variables.
If data transformations are not required, we employ $\mathbf{S}_i$ as is and use the notation $\mathbf{Q}_i = \mathbf{S}_i$ and $m_i = n_i$ for convenience. 

\begin{remark}
In settings in which centering and scaling are needed, we recommend splitting the snapshot dataset by domain region across compute cores, as we did here in the considered example.
This approach enables each core to independently calculate local centering parameters without requiring communication. 
Scaling parameters, which are typically global, can be efficiently determined by first calculating local values on each core followed by a single, inexpensive collective communication step to obtain the global scaling values. 
If centering and scaling are not needed, any non-overlapping splitting scheme of the entire dataset among the $p$ cores suffices.
\end{remark}

\subsubsection{Parallel implementation details}
For the considered 2D Navier-Stokes example, the only employed data transformation is snapshot centering with respect to the temporal mean over the training time horizon.
Since the data splitting was done over the spatial domain, performing centering with respect to the mean is trivial in an MPI program, as illustrated in the code snippet below:
\begin{lstlisting}[language=Python, firstnumber=67]
######## STEP II: DISTRIBUTED DATA TRANSFORMATIONS ########
# compute the temporal mean of each variable on each rank
temporal_mean_rank  = np.mean(Q_rank, axis=1)
# center (in place) each variable with respect to its temporal mean on each rank
Q_rank              -= temporal_mean_rank[:, np.newaxis]
#################### STEP II END ##########################
\end{lstlisting}

Although we do not apply scaling to the centered data in this particular implementation, the following code snippet demonstrates the effectiveness of our data partitioning scheme for computing global maximum absolute values for scaling each variable. 
By distributing the training dataset by domain region, each core can independently calculate local maximum absolute values. 
A subsequent inexpensive global reduction operation determines the overall maximum absolute value for each variable, enabling consistent scaling across all cores.
\begin{lstlisting}[language=Python, numbers=none]
# scale the centered stated variables by their global maximum absolute value
# this ensures that the centered and scaled variables do not exceed [-1, 1]
for j in range(ns):

    # determine the local maximum absolute value of each centered variable on each rank
    min_centered_var_rank   = np.min(Q_rank[j*nx_i : (j + 1)*nx_i, :])
    max_centered_var_rank   = np.max(Q_rank[j*nx_i : (j + 1)*nx_i, :])
    scaling_param_rank 	    = np.maximum(np.abs(min_centered_var_rank), \
                                         np.abs(max_centered_var_rank))

    # determine the global maximum absolute value via a parallel reduction
    # since all ranks require the global scaling parameters, the reduction results are also broadcasted to all ranks
    scaling_param_global = np.zeros_like(scaling_param_rank)
    comm.Allreduce(scaling_param_rank, scaling_param_global, op=MPI.MAX)

    # scale each centered variable by its corresponding global scaling parameter
    Q_rank[j*nx_i : (j + 1)*nx_i, :] /= scaling_param_global
\end{lstlisting} 

\subsection{Step III: Parallel dimensionality reduction}  \label{subsec:dOpInf_step_III}
In the next step, we perform parallel dimensionality reduction, which represent the high-dimensional (transformed) snapshot data with global dimension $m$ in a lower-dimensional space of dimension $r$ such that $r \ll m$.
In our context, the reduced space is a linear subspace, spanned by the column vectors forming the $r$-dimensional POD basis.
However, other approaches such as quadratic manifolds~\cite{GWW23,barnett2022quadratic}, which also support a distributed extension, are viable alternatives.

\subsubsection{Mathematical and computational aspects}

The data dimensionality reduction step is typically the most computationally and memory-intensive part of the standard, serial OpInf formulation. 
This is due to two expensive calculations that depend on the large (transformed) snapshot dimension $m$.
First, the rank-$r$ POD basis is computed, typically via the thin singular value decomposition (SVD)~\cite{GvL96} of $\mathbf{Q}$ at a cost in $\mathcal{O}(mn_t^2)$,
\begin{equation} \label{eq:thin_svd}
    \mathbf{Q} = \mathbf{V} \bm{\Sigma} \mathbf{W}^\top,
\end{equation}
where $\mathbf{V} \in \mathbb{R}^{m \times n_t}$ contains the left singular vectors, $\bm{\Sigma} \in \mathbb{R}^{n_t \times n_t}$ is a diagonal matrix comprising the singular values of $\mathbf{Q}$ in non-decreasing order $\sigma_1 \geq \sigma_2 \geq \ldots \geq \sigma_{n_t}$, where $\sigma_j$ denotes the $j$th singular value, and $\mathbf{W} \in \mathbb{R}^{n_t \times n_t}$ contains the right singular vectors.
The rank-$r$ POD basis $\mathbf{V}_r \in \mathbb{R}^{m \times r}$ is obtained from the first $r$ columns of $\mathbf{V}$ corresponding to the $r$ largest singular values.
This is followed by computing the low-dimensional representation of $\mathbf{Q}$ in the linear subspace spanned by the columns of $\mathbf{V}_r$ as $\hat{\mathbf{Q}} = \mathbf{V}_r^\top \mathbf{Q} \in \mathbb{R}^{r \times n_t}$, at a cost in $\mathcal{O}(r  m  n_t)$.
The cost of the thin SVD can be alleviated by using the randomized~\cite{HMT11} or streaming SVD~\cite{Br06}.
Alternatively, the POD method of snapshots~\cite{BHL93} allows, in principle, handling datasets with large state dimensions by processing subsets of at least two snapshots at a time.
However, its sequential nature and potential for redundancy (e.g., reloading the full dataset for scaling) limit its efficiency.
We will nevertheless show in the following that, starting from the method of snapshots, we can scalably and efficiently represent the high-dimensional snapshot data in the low-dimensional subspace spanned by the $r$-dimensional POD basis vectors without explicitly having to compute the POD basis and without introducing approximations by using, for example, a randomized SVD technique.

Our distributed algorithm starts by computing the Gram matrices $\mathbf{D}_i = \mathbf{Q}_i^\top \mathbf{Q}_i \in \mathbb{R}^{n_t \times n_t}$ on each core, followed by their summation $\mathbf{D} = \sum_{i=1}^p \mathbf{D}_i$ obtained via a parallel reduction. 
Since the original snapshot data was partitioned non-overlappingly, it follows that
\begin{equation} \label{eq:mos_partitioned}
    \mathbf{D} = \sum_{i=1}^p \mathbf{D}_i = \sum_{i=1}^p \mathbf{Q}_i^\top \mathbf{Q}_i = \mathbf{Q}^\top \mathbf{Q}.
\end{equation}
From~\eqref{eq:thin_svd} and~\eqref{eq:mos_partitioned}, we have that
\begin{equation} \label{eq:mos_Sigma_W}
    \mathbf{D} = \mathbf{Q}^\top \mathbf{Q} = \mathbf{W} \mathbf{\Sigma} \mathbf{V}^\top \mathbf{V} \mathbf{\Sigma} \mathbf{W}^\top = \mathbf{W} \mathbf{\Sigma}^2 \mathbf{W}^\top \Rightarrow \mathbf{D} \mathbf{W} = \mathbf{W} \mathbf{\Sigma}^2,
\end{equation}
which implies that the eigenvalues of $\mathbf{D}$ are the squared singular values of $\mathbf{Q}$, and the eigenvectors of $\mathbf{D}$ are equivalent (up to a sign change) to the right singular vectors of $\mathbf{Q}$.
Each core then computes the eigenpairs $\{(\lambda_k, \mathbf{u}_k)\}_{k=1}^{n_t}$ of $\mathbf{D}$, where $\lambda_k$ are the real and non-negative eigenvalues and $\mathbf{u}_k \in \mathbb{R}^{n_t}$ denote the corresponding eigenvectors.
Note that we must ensure that the eigenpairs are arranged such that $\lambda_1 \geq \lambda_2 \geq \ldots \geq \lambda_{n_t}$.

Further, from~\eqref{eq:thin_svd}, it follows that the rank-$r$ POD basis can be computed as
\begin{equation} \label{eq:MoS_POD_basis}
    \mathbf{V}_r = \mathbf{Q} \mathbf{W}_r \bm{\Sigma}_r^{-1} = \mathbf{Q} \mathbf{U}_r \bm{\Lambda}_r^{-\frac{1}{2}}, 
\end{equation}
where $\mathbf{U}_r = \begin{bmatrix} \mathbf{u}_1 \vert \mathbf{u}_2 \vert \dots \vert \mathbf{u}_r \end{bmatrix}$ and $\bm{\Lambda}_r = \mathrm{diag}(\lambda_1, \lambda_2, \ldots, \lambda_r)$.
Equation~\eqref{eq:MoS_POD_basis} implies that
\begin{equation} \label{eq:dOpInf_projection}
    \hat{\mathbf{Q}} = \mathbf{V}_r^\top \mathbf{Q} = \left(\mathbf{Q} \mathbf{U}_r \bm{\Lambda}_r^{-\frac{1}{2}}\right)^\top \mathbf{Q} = \mathbf{T}_r^\top \mathbf{Q}^\top \mathbf{Q} = \mathbf{T}_r^\top \mathbf{D},
\end{equation}
where we used the notation $\mathbf{T}_r =  \mathbf{U}_r \bm{\Lambda}_r^{-\frac{1}{2}} \in \mathbb{R}^{n_t \times r}$. 
Therefore, in our algorithm, the representation of the high-dimensional (transformed) snapshots in the low-dimensional linear subspace spanned by the rank-$r$ POD basis vectors can be efficiently computed in terms of two small matrices, $\mathbf{T}_r$ and $\mathbf{D}$, without explicitly requiring the POD basis.

\subsubsection{Parallel implementation details}

The implementation of the dimensionality reduction step is provided by the code below.
We leverage the \texttt{matmul} function from \texttt{numpy} for computing all (dense) matrix-matrix multiplications, which typically uses highly optimized \texttt{BLAS} routines such as \texttt{dgemm} for double-precision and \texttt{sgemm} for single-precision dense matrix-matrix multiplications.
The eigendecomposition of the symmetric, positive semidefinite global Gram matrix $\mathbf{D}$ is computed via the function \texttt{eigh} from \texttt{numpy}, based on optimized \texttt{LAPACK} routines.
We choose the reduced dimension, $r$, such that the total retained energy corresponding to the first $r$ POD modes is $99.95 \%$, that is,
\begin{equation} \label{eq:ret_energy}
    \frac{\sum_{k=1}^{r} \sigma_k^2}{\sum_{k=1}^{n_t} \sigma_k^2} \geq 0.9995.
\end{equation}
Notice that in our code, the squared singular value ratio in~\eqref{eq:ret_energy} is computed using the eigenvalues of $\mathbf{D}$ (cf.~Eq.~\eqref{eq:mos_Sigma_W}).
In settings where $r$ is prescribed a priori, we need to compute only the first $r$ eigenpairs of $\mathbf{D}$.
This can be done efficiently via the function \texttt{eigsh} from \texttt{scipy}, for example.
Notice also that we compute and broadcast $\mathbf{D}$ to all ranks via the collective MPI function \texttt{Allreduce} (line 79), and all ranks compute the eigendecomposition of $\mathbf{D}$.
This is because the eigenvalues and eigenvalues of $\mathbf{D}$ are used to compute $\mathbf{T}_r$ (line $98$) which is then used to compute $\hat{\mathbf{Q}}$ (line $100$).
$\hat{\mathbf{Q}}$ is required by all ranks for inferring the reduced model operators via OpInf in the next step.
Further, $\mathbf{T}_r$ is also needed by all ranks for postprocessing the reduced solution (see Sec.~\ref{subsec:dOpInf_step_V}).
These computations are cheap when $n_t \ll m$ and $r \ll m$.
An alternative approach would be to compute $\mathbf{D}$ on one rank via a standard reduction, perform all subsequent computations on that rank followed by broadcasting the results to all other ranks. 
However, this strategy requires communication and can create a bottleneck, as all remaining computations must wait for the results, which could reduce the overall efficiency of the code.

\begin{lstlisting}[language=Python, firstnumber=73]
######## STEP III: DISTRIBUTED DIMENSIONALITY REDUCTION ########
# compute the local Gram matrices on each rank
D_rank = np.matmul(Q_rank.T, Q_rank)

# compute the global Gram matrix via a parallel reduction of the local Gram matrices, and broadcast the result to all ranks
D_global = np.zeros_like(D_rank)
comm.Allreduce(D_rank, D_global, op=MPI.SUM)

# compute the eigendecomposition of the positive, semi-definite global Gram matrix
eigs, eigv = np.linalg.eigh(D_global)

# order eigenpairs by increasing eigenvalue magnitude
sorted_indices  = np.argsort(eigs)[::-1]
eigs            = eigs[sorted_indices]
eigv            = eigv[:, sorted_indices]

# define target retained energy for the dOpInf ROM
target_ret_energy = 0.9996

# compute retained energy for r bteween 1 and nt
ret_energy  = np.cumsum(eigs)/np.sum(eigs)
# determine the minimum value of r that exceeds the prescribed retained energy threshold
r           = np.argmax(ret_energy > target_ret_energy) + 1

# compute the auxiliary Tr matrix
Tr_global 	= np.matmul(eigv[:, :r], np.diag(eigs[:r]**(-0.5)))
# compute the low-dimensional representation of the high-dimensional transformed snapshot data
Qhat_global = np.matmul(Tr_global.T, D_global)
##################### STEP III END #############################
\end{lstlisting}

\subsection{Step IV: Parallel reduced operator learning via Operator Inference}  \label{subsec:dOpInf_step_IV}
In the next step, we use distributed computing to infer the reduced operators that define the dOpInf ROM.

\subsubsection{Mathematical and computational aspects}
The Navier-Stokes equations~\eqref{eq:PDE_Navier_Stokes} have only linear and quadratic terms in the governing equations.
We additionally have a constant term introduced into the ROM, due to centering.
The structure-preserving ROM to be inferred therefore takes a quadratic form and reads 
\begin{equation} \label{eq:ROM_quad_time_cont}
    \dot{\hat{\mathbf{q}}} = \bar{\mathbf{A}}\hat{\mathbf{q}} + \bar{\mathbf{H}}\left(\hat{\mathbf{q}} \otimes \hat{\mathbf{q}} \right)  + \bar{\mathbf{c}}.
\end{equation}
OpInf would then proceed by inferring $\bar{\mathbf{c}} \in \mathbb{R}^{r}, \bar{\mathbf{A}} \in \mathbb{R}^{r\times r}$, and $\bar{\mathbf{H}} \in \mathbb{R}^{r\times r^2}$ given the projected snapshot data pairs $\{\dot{\hat{\mathbf{q}}}_k, \hat{\mathbf{q}}_k\}_{k=1}^{n_t}$.
When the high-fidelity code does not provide time derivative data, $\{\dot{\hat{\mathbf{q}}}\}_{k=1}^{n_t}$ must be approximated numerically using $\{\hat{\mathbf{q}}\}_{k=1}^{n_t}$ via finite differences, for example. 
However, such an approximation can be inaccurate, especially when the training snapshots, and consequently $\{\hat{\mathbf{q}}\}_{k=1}^{n_t}$, are temporally downsampled, as it is the case for the considered Navier-Stokes example.
An inaccurate derivative approximation would lead to inaccurate inferred reduced operators and hence inaccurate ROM predictions.
In such situations, we employ the fully discrete formulation of OpInf~\cite{FMW22} that determines the reduced operators $\hat{\mathbf{c}} \in \mathbb{R}^{r}, \hat{\mathbf{A}} \in \mathbb{R}^{r\times r}$, and $\hat{\mathbf{H}} \in \mathbb{R}^{r\times r^2}$ defining the discrete quadratic ROM
\begin{equation} \label{eq:ROM_quad_time_discrete}
    \hat{\mathbf{q}}[k + 1] = \hat{\mathbf{A}}\hat{\mathbf{q}}[k] + \hat{\mathbf{H}}\left(\hat{\mathbf{q}}[k] \otimes \hat{\mathbf{q}}[k] \right) + \hat{\mathbf{c}}
\end{equation}
that best match the projected snapshot data in a minimum residual sense by solving the linear least-squares minimization
\begin{equation} \label{eq:OpInf_reg}
    \mathop{\mathrm{argmin}}_{\hat{\mathbf{O}}} \left\lVert \hat{\mathbf{D}}\hat{\mathbf{O}}^{\top} - \hat{\mathbf{Q}}_2^\top \right\rVert_F^2 + \beta_{1} \left(\left\lVert\hat{\mathbf{A}}\right\rVert_F^2 + \left\lVert\hat{\mathbf{c}}\right\rVert_F^2\right) + \beta_{2} \left\lVert\hat{\mathbf{H}}\right\rVert_F^2,
\end{equation}
where $\hat{\mathbf{O}} =
\begin{bmatrix}
\hat{\mathbf{A}} \, \vert \, \hat{\mathbf{H}} \, \vert \, \hat{\mathbf{c}}
\end{bmatrix}
\in \mathbb{R}^{r \times (r + r^2 + 1)}$ denotes the unknown operators and  
$\hat{\mathbf{D}} =
\begin{bmatrix}
\hat{\mathbf{Q}}_1^\top \, \vert \, \hat{\mathbf{Q}}_1^\top \otimes \hat{\mathbf{Q}}_1^\top \, \vert \, \hat{\mathbf{1}}_{n_t - 1}
\end{bmatrix}
\in \mathbb{R}^{(n_t - 1) \times (r + r^2 +1)}$ the OpInf data, $F$ denotes the Frobenius norm, and
\begin{equation} \label{eq:OpInf_data_matrices_discrete}
    \hat{\mathbf{Q}}_{1} =
     \begin{bmatrix}
\vert & \vert & & \vert\\
     \hat{\mathbf{q}}_1 &
     \hat{\mathbf{q}}_2 &
     \ldots &
     \hat{\mathbf{q}}_{n_t - 1}\\
     \vert & \vert & & \vert
     \end{bmatrix} \in \mathbb{R}^{r \times n_t-1} \quad \text{and} \quad 
     \hat{\mathbf{Q}}_{2} =
     \begin{bmatrix}
\vert & \vert & & \vert\\
     \hat{\mathbf{q}}_2 &
     \hat{\mathbf{q}}_3 &
     \ldots &
     \hat{\mathbf{q}}_{n_t}\\
     \vert & \vert & & \vert
     \end{bmatrix} \in \mathbb{R}^{r \times n_t-1}.
\end{equation}

Equation~\eqref{eq:OpInf_reg} can be efficiently solved using standard least-squares solvers~\cite{GvL96} due to its dependence on the reduced dimension $r$, which is usually small. 
To address overfitting and other sources of error, we introduce regularization hyperparameters $\beta_{1}, \beta_{2} \in \mathbb{R}$. 
Following~\cite{MHW21}, we employ a grid search to find the optimal hyperparameter values. 
This involves exploring a range of candidate values for both $\beta_{1}$ and $\beta_{2} \in \mathbb{R}$ in a nested loop. 
Since the hyperparameters are independent, this search can be easily parallelized. 
The optimal hyperparameters are chosen to minimize the training error, subject to the constraint that the inferred reduced coefficient have bounded growth over a trial time horizon $[t_{\mathrm{init}}, t_{\mathrm{trial}}]$ with $t_{\mathrm{trial}} \geq t_{\mathrm{final}}$~\cite{MHW21, QFW21}.

\subsubsection{Parallel implementation details}
We provide the implementation for the distributed reduced operator learning step below. 
While the code might appear complex, the underlying concepts are straightforward.
Let $\mathcal{B}_1$, $\mathcal{B}_2 \subset \mathbb{R}_{> 0}$ denote the sets of candidate regularization parameters $\beta_1 \in \mathcal{B}_1$ and $\beta_2 \in \mathcal{B}_2$.
For convenience, we choose $\mathcal{B}_1$ and $\mathcal{B}_2$ such that the cardinality of $\mathcal{B}_1 \times \mathcal{B}_2$ is divisible by $p$.
To parallelize the search for the optimal regularization pair, we (i) partition the pairs in 
$\mathcal{B}_1 \times \mathcal{B}_2$ into $p$ subsets, (ii) for each candidate pair, we learn the corresponding reduced operators and compute the reduced solution over $[t_{\mathrm{init}}, t_{\mathrm{trial}}]$, and (iii) determine the rank where the regularization pair that minimizes the training error while ensuring that the inferred reduced coefficients have bounded growth over this trial period resides.

We start with providing the code for four auxiliary functions that we will need later on.
The first function, \texttt{distribute\_reg\_pairs}, splits the the regularization pairs in $\mathcal{B}_1 \times \mathcal{B}_2$ into $p$ subsets.
Next, \texttt{compute\_Qhat\_sq}, computes the unique entries in the quadratic tensor $\hat{\mathbf{Q}}_1^\top \otimes \hat{\mathbf{Q}}_1^\top$ used in the OpInf learning problem~\eqref{eq:OpInf_reg}.
This is because the quadratic reduced operator $\hat{\mathbf{H}} \in \mathbb{R}^{r \times r^2}$ in~\eqref{eq:OpInf_reg} is not uniquely defined as written due to the symmetry of quadratic products (i.e., $\hat{q}_1\hat{q}_2=\hat{q}_2\hat{q}_1$), leading to redundant coefficients.
To ensure a unique solution, we eliminate the redundant degrees of freedom, which amounts to learning an operator of dimension $r \times r(r + 1)/2$ as described in \cite{KPW24}.
The third function computes the training error in the reduced space between the reference projected data $\hat{\mathbf{Q}}$ computed at the end of the dimensionality reduction step and the associated solution computed by the dOpInf ROM, which we denote by $\tilde{\mathbf{Q}} \in \mathbb{R}^{r \times n_t}$. 
The fourth and last function, \texttt{solve\_opinf\_difference\_model}, computes the reduced solution by solving the discrete dOpInf ROM for a specified number of iterations over a target time horizon.
\begin{lstlisting}[language=Python, firstnumber=102]
######## STEP IV: DISTRIBUTED REDUCED OPERATOR INFERENCE ########
def distribute_reg_pairs(rank, n_reg, size):
	"""
 	get_reg_params_per_rank returns the index of the first and last regularization pair for each MPI rank
 
 	:rank:    the MPI rank i = 0, 1, ... , p-1 that will execute this function
 	:n_reg:   total number of regularization parameter pairs
 	:size: 	  size of the MPI communicator (p in our case)

 	:return: the start and end indices of the regularization pairs for each MPI rank
 	"""

	nreg_i_equal = int(n_reg/size)

	start = rank * nreg_i_equal
	end   = (rank + 1) * nreg_i_equal

	if rank == size - 1 and end != n_reg:
		end += n_reg - size*nreg_i_equal

	return start, end

def compute_Qhat_sq(Qhat):
	"""
	compute_Qhat_sq returns the non-redundant terms in Qhat \otimes Qhat

	:Qhat: reduced snapshot data

	:return: the non-redundant in Qhat \otimes Qhat
	"""

	if len(np.shape(Qhat)) == 1:

	    r 		= np.size(Qhat)
	    prods 	= []
	    for i in range(r):
	        temp = Qhat[i]*Qhat[i:]
	        prods.append(temp)

	    Qhat_sq = np.concatenate(tuple(prods))

	elif len(np.shape(Qhat)) == 2:
	    K, r 	= np.shape(Qhat)    
	    prods 	= []
	    
	    for i in range(r):
	        temp = np.transpose(np.broadcast_to(Qhat[:, i], (r - i, K)))*Qhat[:, i:]
	        prods.append(temp)
	    
	    Qhat_sq = np.concatenate(tuple(prods), axis=1)

	else:
	    print('invalid input!')

	return Qhat_sq

def compute_train_err(Qhat_train, Qtilde_train):
	"""
	compute_train_err computes the OpInf training error

	:Qhat_train:   reference data computed by projecting the high-dimensional transformed snapshots
	:Qtilde_train: approximate data computed by the dOpInf ROM

	:return: the value of the training error
	"""
	train_err = \
    np.max(np.sqrt(np.sum( (Qtilde_train - Qhat_train)**2, axis=1) / np.sum(Qhat_train**2, axis=1)))

	return train_err

def solve_discrete_dOpInf_model(qhat0, n_steps_trial, dOpInf_red_model):
	"""
	solve_discrete_dOpInf_model solves the discrete dOpInf ROM for n_steps_trial over a prescribed trial time horizon

	:qhat0:            reduced initial condition
	:n_steps_trial:    number of time steps for OpInf solution
	:dOpInf_red_model: callback to the dOpInf ROM

	:return: a flag indicating NaN presence in the reduced solution and the reduced solution 
	"""

	Qtilde    	    = np.zeros((np.size(qhat0), n_steps_trial))
	contains_nans   = False

	Qtilde[:, 0] = qhat0
	for i in range(n_steps_trial - 1):
	    Qtilde[:, i + 1] = dOpInf_red_model(Qtilde[:, i])

	if np.any(np.isnan(Qtilde)):
	    contains_nans = True
            
	return contains_nans, Qtilde.T
\end{lstlisting}

We now have all the necessary ingredients to determine the optimal regularization parameters in parallel and use them to infer the reduced model operators that define the discrete dOpInf ROM~\ref{eq:ROM_quad_time_discrete}.
We start by prescribing discrete sets of candidate values $\mathcal{B}_1$ and $\mathcal{B}_2$ for the regularization parameters. 
Here, we consider eight candidate values for both $\mathcal{B}_1$ and $\mathcal{B}_2$, but in problems with more complex dynamics, sets with larger cardinalities (and bounds) might be necessary to ensure that the optimal pair leads to accurately inferred reduced operators.
We then prescribe a tolerance for the maximum growth of the inferred reduced coefficients over the trial time horizon, which will be used to determine the optimal regularization pair.
In our implementation, the trial time horizon is the same as the target horizon, that is, $[4, 10]$ seconds.
In the next step, we compute the Cartesian product $\mathcal{B}_1 \times \mathcal{B}_2$ via the function \texttt{product} from \texttt{itertools} followed by distributing the resulting regularization pairs among the $p$ ranks via the function \texttt{distribute\_reg\_pairs}.
After we extract $\hat{\mathbf{Q}}_{1}$ and $\hat{\mathbf{Q}}_{2}$ from $\hat{\mathbf{Q}}$, we compute the non-redundant quadratic terms in  $\hat{\mathbf{Q}}_1^\top \otimes \hat{\mathbf{Q}}_1^\top$ by calling the function \texttt{compute\_Qhat\_sq} (line 223) and assemble the OpInf data matrix $\hat{\mathbf{D}}$ (line 230).

In our implementation, the solution to the OpInf least-squares minimization~\eqref{eq:OpInf_reg} is determined by solving the normal equations~\cite{GvL96}; however, other solvers, such as those based on the SVD or QR decomposition of $\hat{\mathbf{D}}$ can also be used.
To this end, we compute $\hat{\mathbf{D}}^\top \hat{\mathbf{D}}$.
We then determine the maximum deviation from the mean of the reduced training data, which will be needed for selecting the optimal regularization parameter pair.
In the next step, each rank loops over its corresponding regularization pairs, solves the regularized normal equations to determine the constant, linear, and quadratic reduced model operators, and solves the discrete quadratic ROM~\eqref{eq:ROM_quad_time_discrete} for the $1,200$ iterations over the trial time horizon $[4, 10]$ seconds.
We also measure the CPU time of computing the dOpInf ROM solution via the MPI function \texttt{Wtime}.
For all finite solutions, we compute the training error and the maximum absolute growth of the inferred coefficients relative to the maximum absolute growth during training, and filter all solutions for which the maximum growth ratio does not exceed the user-defined tolerance \texttt{max\_growth}.
Finally, we perform an inexepensive collective reduction via MPI \texttt{Allreduce} to find the rank where the optimal regularization pair resides, that is, the pair that both minimizes the training error and satisfies the imposed growth constraint on the inferred reduced coefficients, followed by extracting the corresponding reduced solution and ROM CPU time. 
\begin{lstlisting}[language=Python, firstnumber=194]
# import function to compute the Cartesian product of all candidate regularization pairs
from itertools import product

# ranges for the regularization parameter pairs
B1 = np.logspace(-10., 0., num=8)
B2 = np.logspace(-4., 4., num=8)

# threshold for the maximum growth of the inferred reduced coefficients, used for selecting the optimal regularization parameter pair 
max_growth = 1.2

# compute the Cartesian product of all regularization pairs (beta1, beta2 )
reg_pairs_global    = list(product(B1, B2))
n_reg_global        = len(reg_pairs_global)

# distribute the regularization pairs among the p MPI ranks
start_ind_reg_params, end_ind_reg_params    = \
                        distribute_reg_pairs(rank, n_reg_global, size)
reg_pairs_rank                              = \
                        reg_pairs_global[start_ind_reg_params : end_ind_reg_params]
 
# extract left and right shifted reduced data matrices for the discrete OpInf learning problem
Qhat_1 = Qhat_global.T[:-1, :]
Qhat_2 = Qhat_global.T[1:, :]

# column dimension of the reduced quadratic operator
s = int(r*(r + 1)/2)
# total column dimension of the data matrix Dhat used in the discrete OpInf learning problem
d = r + s + 1

# compute the non-redundant quadratic terms of Qhat_1 \otimes Qhat_1
Qhat_1_sq = compute_Qhat_sq(Qhat_1)

# define the constant part (due to mean shifting) in the discrete OpInf learning problem
K 	 = Qhat_1.shape[0]
Ehat = np.ones((K, 1))

# assemble the data matrix Dhat for the discrete OpInf learning problem
Dhat   = np.concatenate((Qhat_1, Qhat_1_sq, Ehat), axis=1)
# compute Dhat.T @ Dhat for the normal equations to solve the OpInf least squares minimization
Dhat_2 = Dhat.T @ Dhat

# compute the temporal mean and maximum deviation of the reduced training data
mean_Qhat_train     = np.mean(Qhat_global.T, axis=0)
max_diff_Qhat_train = np.max(np.abs(Qhat_global.T - mean_Qhat_train), axis=0)

# dictionary to store the regularization pair for each regularization pair
opt_train_err_reg_pair      = {}
# dictionary to store the reduced solutions over the target time horizon for each pair
Qtilde_dOpInf_reg_pair      = {}
# dictionary to store the OpInf ROM CPU time for each regularization pair
dOpInf_ROM_rtime_reg_pair   = {}

# loop over the regularization pairs corresponding to each MPI rank
for pair in reg_pairs_rank:

    # extract beta1 and beta2 from each candidate regularization pair
    beta1 = pair[0]
    beta2 = pair[1]

    # regularize the linear and constant reduced operators using beta1, and the reduced quadratic operator using beta2
    regg            = np.zeros(d)
    regg[:r]        = beta1
    regg[r : r + s] = beta2
    regg[r + s:]    = beta1
    regularizer     = np.diag(regg)
    Dhat_2_reg      = Dhat_2 + regularizer

    # solve the OpInf learning problem by solving the regularized normal equations
    Ohat = np.linalg.solve(Dhat_2_reg, np.dot(Dhat.T, Qhat_2)).T

    # extract the linear, quadratic, and constant reduced model operators
    Ahat = Ohat[:, :r]
    Fhat = Ohat[:, r:r + s]
    chat = Ohat[:, r + s]

    # define the discrete dOpInf reduced model 
    dOpInf_red_model    = lambda x: Ahat @ x + Fhat @ compute_Qhat_sq(x) + chat
    # extract the reduced initial condition from Qhat_1
    qhat0               = Qhat_1[0, :]
    
    # compute the reduced solution over the trial time horizon, which here is the same as the target time horizon
    start_time_dOpInf_eval          = MPI.Wtime()
    contains_nans, Qtilde_dOpInf    = solve_discrete_dOpInf_model(qhat0, nt_p, dOpInf_red_model)
    end_time_dOpInf_eval            = MPI.Wtime()

    time_dOpInf_eval = end_time_dOpInf_eval - start_time_dOpInf_eval

    # for each candidate regularization pair, we compute the training error 
    # we also save the corresponding reduced solution and ROM evaluation time
    # and compute the ratio of maximum coefficient growth in the trial period to that in the training period
    opt_train_err                               = 1e20
    opt_train_err_reg_pair[opt_train_err]       = pair
    Qtilde_dOpInf_reg_pair[opt_train_err]       = Qtilde_dOpInf
    dOpInf_ROM_rtime_reg_pair[opt_train_err]    = time_dOpInf_eval

    if contains_nans == False:
        train_err               = compute_train_err(Qhat_global.T[:nt, :], Qtilde_dOpInf[:nt, :])
        max_diff_Qhat_trial     = np.max(np.abs(Qtilde_dOpInf - mean_Qhat_train), axis=0)		
        max_growth_trial        = np.max(max_diff_Qhat_trial)/np.max(max_diff_Qhat_train)

        if max_growth_trial < max_growth:
            opt_train_err                               = train_err
            opt_train_err_reg_pair[opt_train_err]       = pair
            Qtilde_dOpInf_reg_pair[opt_train_err]       = Qtilde_dOpInf
            dOpInf_ROM_rtime_reg_pair[opt_train_err]    = time_dOpInf_eval

# find the globally minimum training error by reducing local results on each rank, subject to the bound constraint on the inferred reduced coefficients over the prescribed trial time horizon
opt_key_rank    = np.min(list(opt_train_err_reg_pair.keys()))
opt_key_rank    = np.array([opt_key_rank])
opt_key_global  = np.zeros_like(opt_key_rank)

comm.Allreduce(opt_key_rank, opt_key_global, op=MPI.MIN)
opt_key_global = opt_key_global[0]

# extract the optimal regularization pair, and the corresponding reduced solution and dOpInf ROM CPU time
if opt_key_rank == opt_key_global:
    rank_reg_opt        = rank
    Qtilde_dOpInf_opt   = Qtilde_dOpInf_reg_pair[opt_key_global]

    reg_pair_opt = opt_train_err_reg_pair[opt_key_global]

    beta1_opt = reg_pair_opt[0]
    beta2_opt = reg_pair_opt[1]

    dOpInf_ROM_rtime_opt = dOpInf_ROM_rtime_reg_pair[opt_key_global]
else:
    rank_reg_opt        = -1
    Qtilde_dOpInf_opt   = None
####################### STEP IV END #############################
\end{lstlisting}

\subsection{Step V: Parallel postprocessing of the reduced solution}  \label{subsec:dOpInf_step_V}
In the final step, we show how to postprocess the obtained reduced solution in parallel. 

\subsubsection{Mathematical and computational aspects}
Postprocessing the reduced solution generally involves mapping it back to the original high-dimensional coordinates for visualization purposes, or for computing errors or other metrics for assessing accuracy.
Let $\Tilde{\mathbf{Q}} \in \mathbb{R}^{r \times n_p}$ denote the matrix containing the reduced solution obtained at the end of the previous step.
To map $\Tilde{\mathbf{Q}}$ to the original coordinates, we first need to multiply $\Tilde{\mathbf{Q}}$ by the components of the rank-$r$ POD basis  $\mathbf{V}_{r, i} \in \mathbb{R}^{m_i \times r}$ to obtain the high-dimensional representation of the solution on each rank followed by applying any inverse data transformations to obtain the approximate solution in the same coordinates as the reference solution.
To this end, we need to first broadcast $\Tilde{\mathbf{Q}}$ from the MPI rank that computed it (i.e., the rank that contains the optimal regularization pair) to all other ranks.
We then use Eq.~\eqref{eq:MoS_POD_basis}, stemming from the method of snapshots, to compute the components of the rank-$r$ POD basis $\mathbf{V}_{r, i}$ on each rank.
We then lift $\Tilde{\mathbf{Q}}$ to the original high-dimensional space by computing independent matrix-matrix products $\mathbf{V}_{r, i} \Tilde{\mathbf{Q}} \in \mathbb{R}^{m_i \times n_t}$, followed by applying any necessary inverse data transformations to the lifted high-dimensional components on each core.
Neither of these two steps require communication.
The result of these operations is the high-dimensional approximate ROM solution over the target time horizon.
However, in scenarios where only certain spatial locations or time instants are of interest, the postprocessing cost can be reduced by computing only the necessary components of the target approximate solution.

\subsubsection{Parallel implementation details}
The following code demonstrates how to use the reduced solution $\Tilde{\mathbf{Q}} \in \mathbb{R}^{r \times 1,200}$ to compute the two approximate velocity components in the original coordinates at three probe locations positioned near the mid-channel, with increasing distance from the circular cylinder, namely $(0.40, 0.20), (0.60, 0.20)$, and $(1.00, 0.20)$.
The corresponding grid point indices within each snapshot for these locations are $\{16,992; 48,250; 130,722\}$; we provide a script in the repository that extracts grid point indices for user-specified probe locations.
We start by broadcasting $\Tilde{\mathbf{Q}} \in \mathbb{R}^{r \times 1,200}$ from the rank holding the optimal regularization pair to all other ranks.
We then map the global probe indices to local indices to identify which rank contains the respective solutions, compute the corresponding components of the local rank-$r$ POD bases, map the reduced solution for each probe to the original coordinates, and save the approximate solutions to disk.
Overall, this postprocessing step is computationally inexpensive as it targets only specific probe locations.
\begin{lstlisting}[language=Python, firstnumber=323]
######## POSTPROCESSING ########
# indices for probe locations (0.40, 0.20), (0.60, 0.20) and (1.00, 0.20)
target_probe_indices = [48250, 77502, 130722]

# broadcast the dOpInf reduced solution from the rank having the optimal regularization pair to all ranks
if rank == rank_reg_opt:
    for i in range(size):
        if i != rank_reg_opt:
            comm.send(Qtilde_dOpInf_opt, dest=i)
else:
    Qtilde_dOpInf_opt = comm.recv(Qtilde_dOpInf_opt, source=rank_reg_opt)

# extract and save to disk the approximate solutions at the probe locations with indices specified in target_probe_indices
for target_var_index in range(ns):
    for j, probe_index in enumerate(target_probe_indices):

        # map the global probe indices to local indices
        if probe_index >= nx_i_start and probe_index < nx_i_end:
            probe_index -= nx_i_start			

            # extract the components of the POD basis corresponding to the probe indices
            Phir_probe 	        = \
                    np.matmul(Q_rank[probe_index + target_var_index*nx_i, :], Tr_global)
            # do the same for the temporal mean used for centering
            temporal_mean_probe = \
                    temporal_mean_rank[probe_index + target_var_index*nx_i]

            # map the reduced solution for each probe location to the original coordinates
            var_probe_prediction = Phir_probe @ Qtilde_dOpInf_opt.T + temporal_mean_probe

            np.save('postprocessing/dOpInf_postprocessing/dOpInf_probe_' + str(j + 1) + \
                '_var_' + str(target_var_index + 1) + '.npy', var_probe_prediction)
###### POSTPROCESSING END ######
\end{lstlisting}

\section{Results obtained with the distributed algorithm in the Navier-Stokes example}\label{sec:results}

\begin{figure}[htp]
\centering
\includegraphics[width=1.0\textwidth]{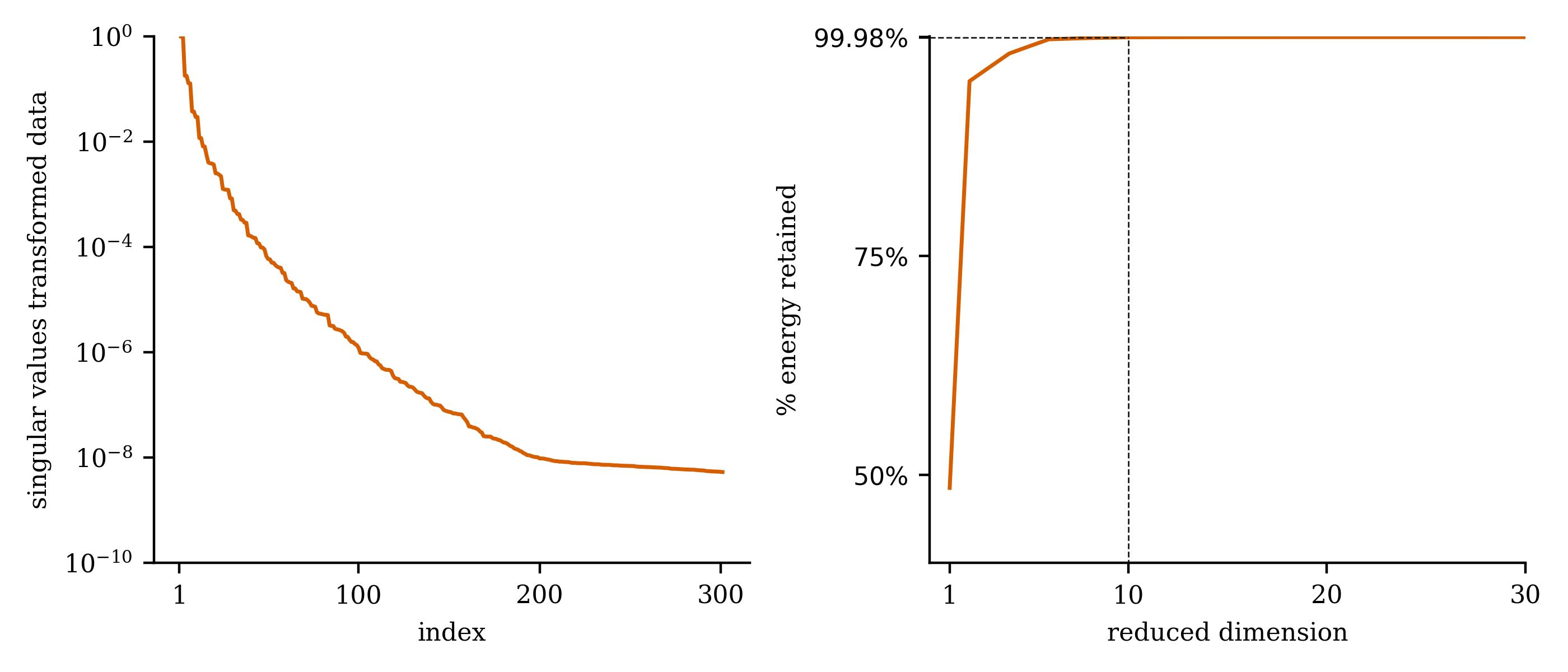}
\caption{Normalized singular values (left) and corresponding retained energy (right) for the considered 2D Navier-Stokes example.}
\label{fig:singular_values}
\end{figure}

This section briefly summarizes the results obtained using the dOpInf algorithm.
Figure~\ref{fig:singular_values} plots the normalized singular values (on the left) and corresponding retained energy (on the right) obtained with our distributed algorithm.
Since the singular values decay fast, $r=10$ POD modes suffice to attain the prescribed energy threshold of $99.96 \%$.
The reduced dimension of the target dOpInf ROM is therefore $r=10$.

\begin{figure}[htp]
\centering
\includegraphics[width=1.0\textwidth]{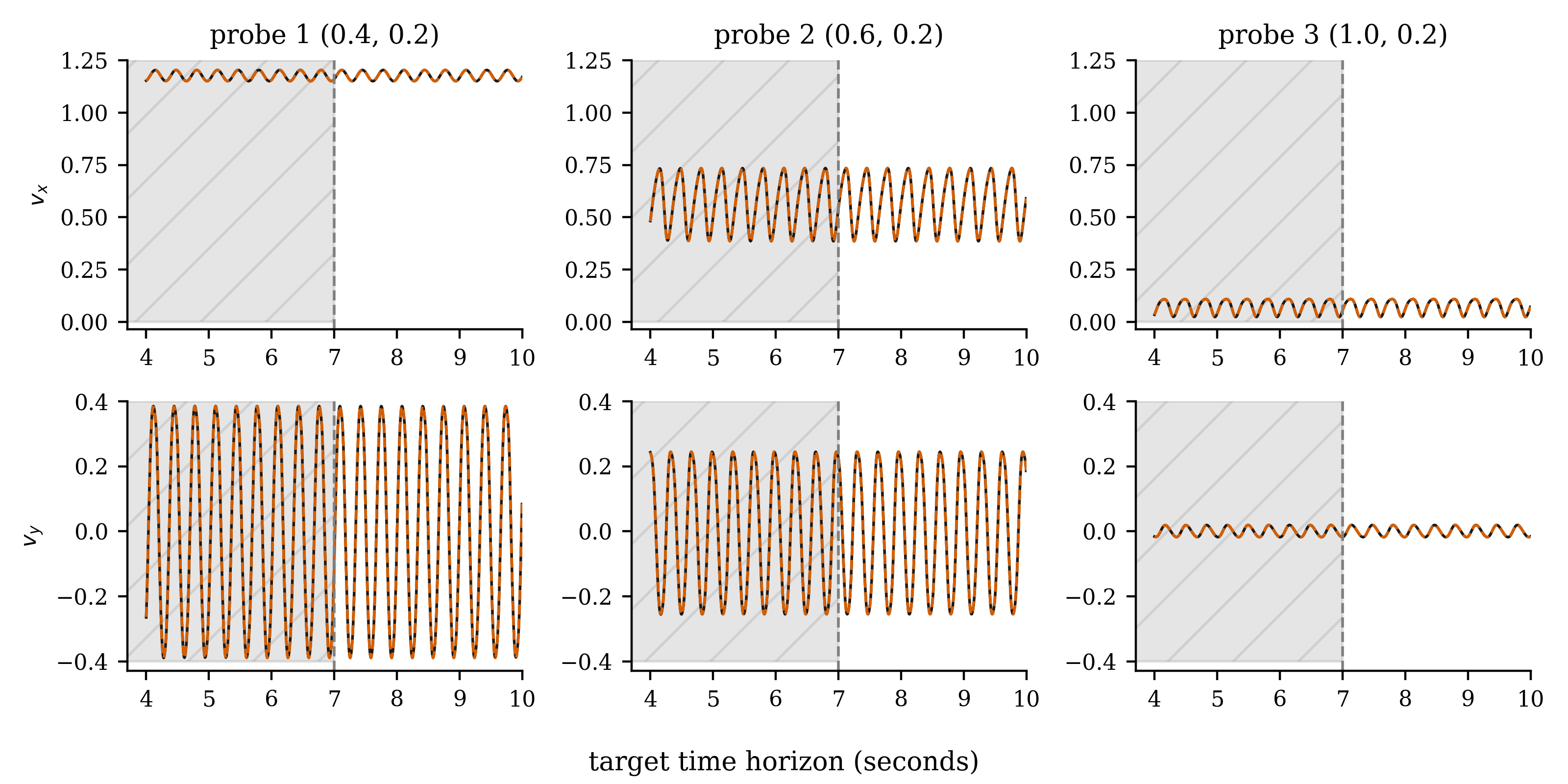}
\caption{Approximate velocity solutions at three probe locations. The hashed areas mark the training horizon.}
\label{fig:probes}
\end{figure}

The optimal regularization pair found by our algorithm is $(\beta_1^* = 7.19\times 10^{-8}, \beta_2^* = 51.79)$.
This results in an accurate reduced model, as demonstrated by Fig.~\ref{fig:probes}, which plots the ROM approximate solutions at the three probe locations, $(0.40, 0.20), (0.60, 0.20)$, and $(1.00, 0.20)$.
The hashed areas on the left mark the training time horizon.

Finally, we present a brief analysis of the CPU time required by the dOpInf algorithm when applied to the 2D Navier-Stokes example.
The goal of this analysis is to provide users with further details into the algorithm rather than delving into detailed scalability considerations.
For a discussion of the scalability of the dOpInf algorithm in large-scale applications, we refer the reader to Ref.~\cite{farcas2024distributedcomputingphysicsbaseddatadriven}.

\begin{figure}[htp]
\centering
\includegraphics[width=1.0\textwidth]{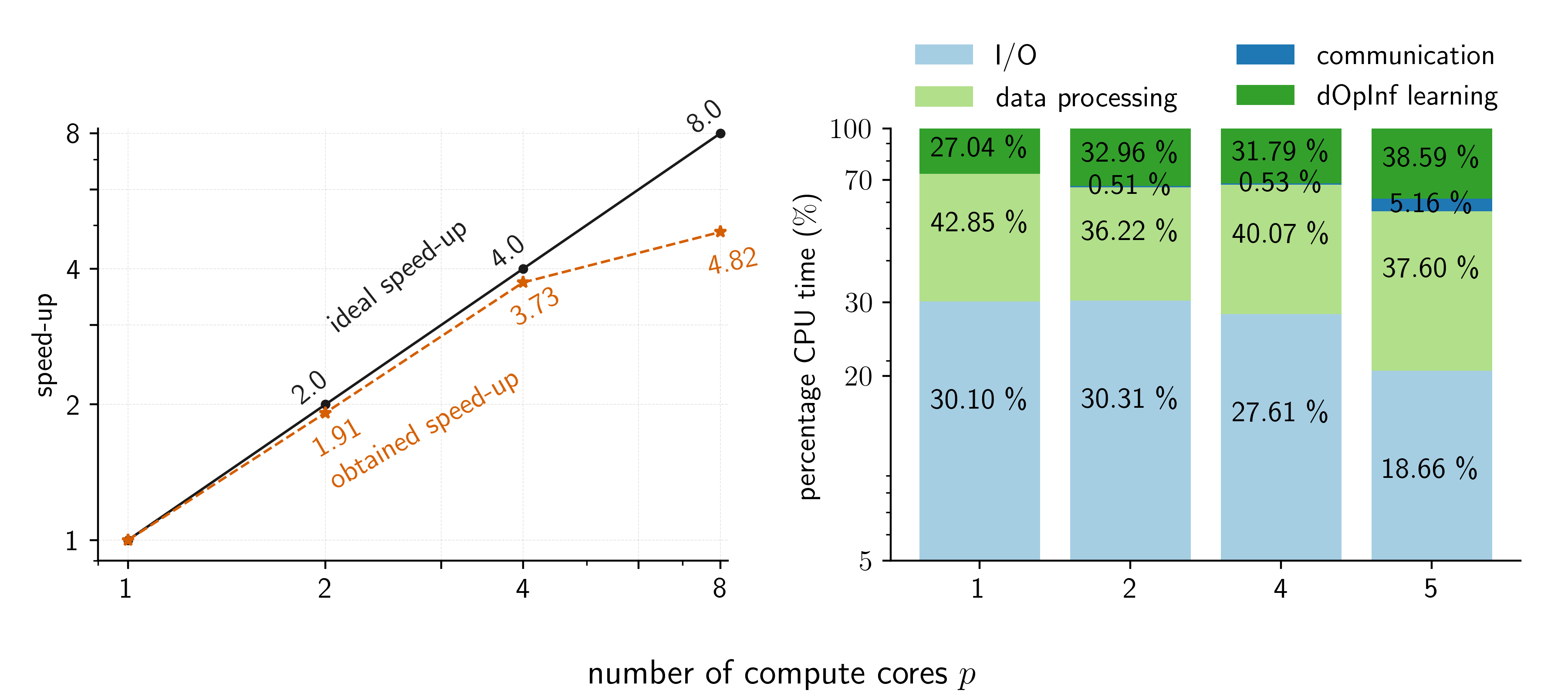}
\caption{Assessing strong scalability for $p \in \{1, 2, 4, 8\}$ cores for the 2D Navier-Stokes example.}
\label{fig:scalability}
\end{figure}

We measure the CPU time of running Steps I--IV (i.e., from loading the training data to computing the reduced solution) using $p \in \{1, 2, 4, 8\}$ compute cores on a shared-memory machine with $256$ AMD EPYC 7702 CPUs and $2$ TB of RAM.
For $p=1$, we run the corresponding serial implementation of OpInf, which is also provided in the repository.
To decrease the effect of fluctuations, we performed each measurement $100$ times.
For dOpInf, we record the CPU time of the MPI rank that contains the optimal regularization parameter pair.
Figure~\ref{fig:scalability} plots the results.
On the left, we plot the speed-up obtained using the mean CPU times as $p$ increases.
For reference, the CPU times are $8.35 \pm 0.40$ seconds for $p=1$; $4.35 \pm 0.02$ seconds for $p=2$; $2.23 \pm 0.09$ seconds for $p=4$; $1.72 \pm 0.18$ seconds for $p=8$.
Moreover, the CPU time of the dOpInf ROM is $0.03 \pm 0.002$ seconds.
The speed-up is excellent up to $p=4$ cores, but it starts to deteriorate for $p=8$.
As the number of cores $p$ increases, the proportion of the code that can be parallelized (namely, computations involving distributed snapshot data) diminishes in this simple example, resulting in a more significant serial component.
This trend is illustrated in Fig.~\ref{fig:scalability} on the right, which presents a breakdown of the CPU time into data loading, all data processing computations, communication, and dOpInf learning.
The communication overhead as $p$ grows contributes to the deterioration in the speed-up.
Note that the deterioration in speed-up here is due to the relatively small size of the example; In Ref.~\cite{farcas2024distributedcomputingphysicsbaseddatadriven}, near-ideal speed-ups were observed for up to p=$2,048$ cores.

\section{Summary and conclusions} \label{sec:conclusion}
Distributed Operator Inference represents a tool for researchers and engineers working on large-scale aerospace engineering simulations, enabling the efficient and scalable construction of computationally efficient physics-based data-driven reduced models.
Its key characteristics as follows.
\emph{Scalability}: Distributed Operator Inference is fully parallelizable and can effectively utilize leadership high-performance computing platforms, making it suitable for handling datasets with large state dimensions. 
\emph{Efficiency}: This algorithm efficiently processes and learns structured reduced models from large-scale datasets, enabling rapid modeling of complex systems. 
\emph{Computational efficiency}: The resulting reduced models are computationally inexpensive, making them ideal for design exploration and risk assessment.
The paper provided a comprehensive description of the algorithm and a step-by-step implementation in a two-dimensional benchmark example, guiding users on practical implementation details and facilitating seamless integration with complex aerospace engineering applications.
The full tutorial is available at \url{https://github.com/ionutfarcas/distributed_Operator_Inference}.

\section*{Acknowledgements}
This work was supported in part by AFRL Grant FA9300-22-1-0001 and the Air Force Center of Excellence on Multifidelity Modeling of Rocket Combustor Dynamics under grant FA9550-17-1-0195.
The views expressed are those of the author and do not necessarily reflect the official policy or position of the Department of the Air Force, the Department of Defense, or the U.S. government.
We thank Dr.~Rudy Geelen for useful discussion regarding the implementation of the considered Navier-Stokes benchmark example.

\bibliography{dOpInf_tutorial}
\end{document}